\documentclass[aps,prd,twocolumn,nofootinbib]{revtex4-2}

\usepackage{graphicx}
\usepackage{amsmath}
\usepackage{amssymb}
\usepackage{xcolor}
\usepackage{extpfeil}
\usepackage{nicematrix}
\usepackage{setspace}
\usepackage{CJK}
\usepackage[colorlinks,citecolor=blue,linkcolor=blue,urlcolor=blue,anchorcolor=blue]{hyperref}
\allowdisplaybreaks
\newcommand\dx{{\rm d}}

\newcommand\AJ{Astron. J.}

\newcommand\ApJ{Astrophys. J.}
\newcommand\ApJL{Astrophys. J. Lett.}
\newcommand\ApJS{Astrophys. J. Suppl.}
\newcommand\ARAA{Annu. Rev. Astron. Astrophys.}
\newcommand\ARNPS{Annu. Rev. Nucl. Part. Sci.}
\newcommand\AstronAstrophys{Astron. Astrophys.}

\newcommand\ChinPhys{Chin. Phys.}
\newcommand\ChinPhysC{Chin. Phys. C}

\newcommand\ComputPhysCommun{Comput. Phys. Commun.}

\newcommand\JCAP{J. Cosmol. Astropart. Phys.}

\newcommand\LivingRevRelativity{Living Rev. Relativity}
\newcommand\MNRAS{Mon. Not. R. Astron. Soc.}

\newcommand\ModPhysLettA{Mod. Phys. Lett. A}

\newcommand\NatRevPhys{Nat. Rev. Phys.}
\newcommand\Nature{Nature (London)}

\newcommand\NuclPhysB{Nucl. Phys. B}

\newcommand\PhysDarkUniverse{Phys. Dark Universe}
\newcommand\PhysRep{Phys. Rep.}
\newcommand\PhysRev{Phys. Rev.}
\newcommand\PhysRevResearch{Phys. Rev. Research}
\newcommand\PLB{Phys. Lett. B}

\newcommand\PRD{Phys. Rev. D}
\newcommand\PRL{Phys. Rev. Lett.}

\newcommand\RMP{Rev. Mod. Phys.}

\begin{document}

\begin{CJK*}{UTF8}{gbsn}
\title{Cosmological consequences of a scalar field with oscillating equation of state. IV. Primordial nucleosynthesis and the deuterium problem}
\author{S. X. Tian (田树旬)}
\email[]{tshuxun@bnu.edu.cn}
\affiliation{Department of Astronomy, Beijing Normal University, 100875 Beijing, China}
\date{\today}
\begin{abstract}
  We study the primordial nucleosynthesis (BBN) in the stepwise scalar field model proposed by Ti\'an [\href{https://doi.org/10.1103/PhysRevD.101.063531}{Phys. Rev. D {\bf 101}, 063531 (2020)}], which provides a multiaccelerating Universe solution to the cosmological coincidence problem and predicts that the scalar field may be non-negligible even in the early Universe. The observed abundances of the light elements can be used to constrain the energy density of the scalar field during the BBN era. We present a public \texttt{Matlab} code to implement the BBN calculation in the stepwise scalar field model. We show that the model can survive the BBN constraints. In particular, this model incorporates a new solution to the possible deuterium problem: very early dark energy that appears at the end of BBN. In addition, the BBN constraints, along with constraints from the cosmic late-time acceleration, suggest that the Universe in the radiation era evolves in a chaotic accelerating manner, rather than an oscillating scaling manner.
\end{abstract}
\maketitle
\end{CJK*}

\section{Introduction}\label{sec:01}
The current standard cosmological model holds that there are two stages of accelerating expansion in the Universe \cite{Frieman2008.ARAA.46.385,Bartelmann2010.RMP.82.331}. One is the early inflation, which is proposed to solve the horizon and flatness problems \cite{Guth1981.PRD.23.347,Linde1982.PLB.108.389,Albrecht1982.PRL.48.1220}. The other is the cosmic late-time acceleration, which was confirmed by observations of supernovae \cite{Riess1998.AJ.116.1009,Perlmutter1999.ApJ.517.565}. Theoretically, there are many dark energy models to explain the late-time acceleration, such as the cosmological constant (the standard model), quintessence \cite{Peebles1988.ApJL.325.L17,Caldwell1998.PRL.80.1582,Steinhardt1999.PRD.59.123504,Zlatev1999.PRL.82.896}, phantom \cite{Caldwell2002.PLB.545.23}, $k$-essence \cite{Armendariz-Picon2000.PRL.85.4438,Armendariz-Picon2001.PRD.63.103510}, quintom \cite{Feng2005.PLB.607.35,Guo2005.PLB.608.177,Cai2010.PhysRep.493.1}. However, these models share a common problem --- the cosmological coincidence problem, which states why the energy density of dark energy is now on the same order of magnitude as that of matter \cite{Carroll2001.LivingRevRelativity.4.1,Bull2016.PhysDarkUniverse.12.56}. To solve this problem, \citet{Dodelson2000.PRL.85.5276} proposed that the Universe may have experienced periodic accelerating expansion in the past and the current acceleration is just a natural continuation. Meanwhile they implemented this scenario with a canonical scalar field model. This multiacceleration scenario inspired many follow-up studies, including model extensions \cite{Griest2002.PRD.66.123501,Ahmed2004.PRD.69.103523,Feng2006.PLB.634.101,Nojiri2006.PLB.637.139,Zhao2007.ChinPhys.16.2830} and confronting observations \cite{Xia2005.ModPhysLettA.20.2409,Xia2006.PRD.74.083521,Pace2012.MNRAS.422.1186,Pan2018.PRD.98.063510}. In particular, inflation can be naturally unified into this scenario \cite{Griest2002.PRD.66.123501,Feng2006.PLB.634.101,Nojiri2006.PLB.637.139}, which makes the model more attractive. We proposed a new canonical scalar field model to realize the multiacceleration scenario \cite{Tian2020.PRD.101.063531}. Hereafter we call this type of model a stepwise scalar field model since the potential is similar to the staircase (see Fig. 1 in \cite{Tian2020.PRD.101.063531} for an illustration). Furthermore, we found chaos and period-doubling bifurcations in our model \cite{Tian2020.PRD.102.063509}, which were not identified in previous models. Our model also naturally unifies inflation and predicts an extremely small tensor-to-scalar ratio \cite{Tian2021.PRD.103.123545}, which is consistent with current observations \cite{Ade2021.PRL.127.151301}.

As the literal meaning of multiacceleration expresses, there should also be stages of accelerating expansion in the early Universe, that is, stages dominated by dark energy. How to detect dark energy in the early Universe? One approach is the global cosmological parameter constraints as did in \cite{Xia2005.ModPhysLettA.20.2409,Xia2006.PRD.74.083521,Pan2018.PRD.98.063510}. This approach has strong model dependencies and thus is indirect. Another approach is the primordial nucleosynthesis (Big Bang nucleosynthesis, abbreviated BBN, see \cite{Cyburt2016.RMP.88.015004,Pitrou2018.PhysRep.754.1} for reviews), which could probe the Universe with temperatures between $10^{12}\,{\rm K}$ and $10^{8}\,{\rm K}$. Given the great success (mainly about the helium-4 and deuterium abundances) of the standard BBN theory at the time, \citet{Dodelson2000.PRL.85.5276} attempted to completely hide the scalar field during the BBN era in their model. Here standard means there is no dark energy. There are other similar works using BBN to constrain the upper limit of the early dark energy density in specific models \cite{Bean2001.PRD.64.103508,Arbey2019.JCAP.11.038}. However, with the improvement of the measurement uncertainties about nuclear reaction rates and cosmological parameters, the standard BBN theory now seems to be broken. In addition to the well-known primordial lithium problem \cite{Fields2011.ARNPS.61.47}, there are inconsistencies between the theoretical prediction and observational abundance about deuterium \cite{Pitrou2021.MNRAS.502.2474,Pitrou2021.NatRevPhys.3.231}. The essence of this deuterium problem remains unclear. On the one hand, it is still debatable whether the deuterium problem really exists \cite{Yeh2021.JCAP.03.046,Pisanti2021.JCAP.04.020}. On the other hand, if the deuterium problem does exist, its solution may hint at new physics, such as time-varying fundamental constants \cite{Deal2021.AstronAstrophys.653.A48}. The deuterium and lithium problems may also shed light on the possible existence of dark energy in the early Universe.

In this paper, we use BBN to test the stepwise scalar field model proposed by us \cite{Tian2020.PRD.101.063531}. We focus on the following two issues:
\begin{enumerate}
  \item Is it possible to completely hide the scalar field in our model as did in \cite{Dodelson2000.PRL.85.5276}?
  \item Is there some kind of early dark energy profiles in our model to solve the deuterium problem, or even the lithium problem?
\end{enumerate}
Meanwhile we require that the model under the same parameter settings can explain the observed cosmic late-time acceleration. In this paper, we intend to provide case studies rather than global parameter constraints.

This paper is organized as follows. Section \ref{sec:02} summarizes the BBN theory, including how to incorporate the stepwise scalar field into the BBN calculation. Section \ref{sec:03} describes our numerical results, including discussions about the BBN results of the standard model, and models with exponential scalar field and stepwise scalar field. Conclusions are presented in Sec. \ref{sec:04}. Throughout this paper, we adopt the SI Units and retain all physical constants unless otherwise stated. All reported uncertainties represent $68\%$ confidence intervals.

\section{Basic theory}\label{sec:02}
BBN calculation requires one to solve a nuclear reaction network in the expanding Universe at temperatures between $10^{12}\,{\rm K}$ and $10^{8}\,{\rm K}$ (equivalently energy scales between $100\,{\rm MeV}$ and $0.01\,{\rm MeV}$). The main BBN physics was summarized in \cite{Wagoner1967.ApJ.148.3,Wagoner1969.ApJS.18.247}, and the results are widely used in current BBN codes, e.g., \texttt{NUC123} \cite{Kawano1988.FermiReport,Kawano1992.FermiReport}, \texttt{PArthENoPE} \cite{Pisanti2008.ComputPhysCommun.178.956,Consiglio2018.ComputPhysCommun.233.237,Gariazzo2022.ComputPhysCommun.271.108205}, \texttt{AlterBBN} \cite{Arbey2012.ComputPhysCommun.183.1822,Arbey2020.ComputPhysCommun.248.106982} and \texttt{PRIMAT} \cite{Pitrou2018.PhysRep.754.1,Pitrou2021.MNRAS.502.2474}. Here we summarize the BBN theory implemented in this paper, which include all the key physics. The results are integrated into a public \texttt{Matlab} code, which is named \texttt{BBNLab} and is available at GitHub\footnote{\url{https://github.com/tshuxun/BBNLab}}. Note that no physics in this section is original. However, this summary, especially the explicit expressions given in Eqs. (\ref{eq:13}) and (\ref{eq:15}), may be useful for beginners and presents a clear basis for our future BBN work.

\subsection{Evolution equations}\label{sec:0201}
The Universe is assumed to be described by the flat Friedmann-Lema\^itre-Robertson-Walker metric $\dx s^2=-c^2\dx t^2+a^2(\dx x^2+\dx y^2+\dx z^2)$, where $a=a(t)$ is the cosmic scale factor and $c$ is the speed of light, and contains photons, neutrinos, electrons, positrons, baryons, and a canonical scalar field. Hereafter we use the subscripts $\{\gamma,\nu,e^-,e^+,b,\phi\}$ to denote these ingredients respectively, use plasma to refer \{photons + electrons + positrons + baryons\}, use the subscript matter to refer \{plasma + neutrinos\}, and use the subscript total to refer \{matter + $\phi$\}. During BBN era, we assume that the plasma is in thermal equilibrium, while the neutrino gas is thermally decoupled from the plasma\footnote{Note that neutrinos are in thermal equilibrium with the plasma through weak interactions when the energy scale is higher than $2\,{\rm MeV}$ \cite{Pitrou2018.PhysRep.754.1}. As the temperature drops, neutrino decoupling and $e^\pm$ annihilation occur one after the other. Neither process is instantaneous and there is a temporal overlap between them. This overlap results in a neutrino heating process from the $e^\pm$ annihilation \cite{Mangano2005.NuclPhysB.729.221}. For simplicity, we assume that all the influences of this neutrino heating process on BBN are reflected in a constant $N_\nu^{\rm eff}$ [see Eq. (\ref{eq:02b})]. After considering $N_\nu^{\rm eff}$, the neutrino gas can be regarded as expanding adiabatically. In addition, the scalar field that appears during the $e^\pm$ annihilation may slightly change the value of $N_\nu^{\rm eff}$ as it affects the Hubble expansion rate. In this paper, we ignore this possible modification and would like to leave this issue to the future.}. The scalar field interacts with other species only through gravity. Nuclides considered in the nuclear network and their basic properties are listed in Table \ref{tab:01}. Substituting the metric into the Einstein field equations gives the Friedmann equation
\begin{equation}\label{eq:01}
  H^2=\frac{8\pi G}{3}(\rho_\gamma+\rho_\nu+\rho_e+\rho_b+\rho_\phi),
\end{equation}
where $H\equiv\dot{a}/a$ is the Hubble parameter, $\dot{}\equiv\dx/\dx t$, $G$ is the Newtonian constant of gravitation, $\rho_e=\rho_{e^-}+\rho_{e^+}$. The mass densities (${\rm energy\ densities}/c^2$) are given by
\begin{subequations}
\begin{align}
  \rho_\gamma &= \frac{\pi^2k_{\rm B}^4T_\gamma^4}{15\hbar^3c^5}, \label{eq:02a}\\
  \rho_\nu    &= \frac{7N_\nu^{\rm eff}\pi^2k_{\rm B}^4T_\nu^4}{120\hbar^3c^5}, \label{eq:02b}\\
  \rho_e      &= \frac{k_{\rm B}^4T_\gamma^4}{\pi^2\hbar^3c^5}\int_{z_\gamma}^{\infty}x^2(x^2-z_\gamma^2)^{1/2}\times \nonumber\\
              &  \qquad\qquad\qquad\left(\frac{1}{e^{x-\phi_e}+1}+\frac{1}{e^{x+\phi_e}+1}\right)\dx x, \nonumber\\
              &= \frac{2m_e^4c^3}{\pi^2\hbar^3}\sum_{i=1}^{\infty}(-1)^{i+1}\cosh(i\phi_e)M(iz_\gamma), \label{eq:02c}\\
  \rho_b      &= n_b\sum_{i=1}^{N_{\rm nuclei}}Y_i(m_i+\frac{3m_e}{2z_\gamma}), \label{eq:02d}\\
  \rho_\phi   &= \frac{c^2}{8\pi G}\left[\frac{\dot{\phi}^2}{2c^2}+V(\phi)\right], \label{eq:02e}
\end{align}
\end{subequations}
where $k_{\rm B}$ is the Boltzmann constant, $\hbar$ is the reduced Planck constant, $T_\gamma$ is the plasma temperature, $T_\nu$ is the neutrino temperature, $N_\nu^{\rm eff}=3.046$ is the effective number of neutrinos \cite{Mangano2005.NuclPhysB.729.221}, $\infty$ means positive infinity, $z_\gamma\equiv m_ec^2/(k_{\rm B}T_\gamma)$, $\phi_e\equiv\mu_e/(k_{\rm B}T_\gamma)$, $m_e$ is the mass of electron, $\mu_e$ is the chemical potential of the electron gas, $M(x)$ is a special function defined in Appendix \ref{App:A}, $m_i$ is the mass of nuclide $i$ (see Table \ref{tab:01}), $Y_i\equiv n_i/n_b$, $n_i$ is the number density of nuclide $i$, $n_b$ is the total number density of baryons, $N_{\rm nuclei}$ is the total species of nuclides, $V(\phi)$ is the potential of the scalar field (see Sec. \ref{sec:0201a}). The conventions about the scalar field and also $\rho_\phi$ are consistent with \cite{Tian2020.PRD.101.063531}. The black-body spectrum is used to derive Eq. (\ref{eq:02a}). The Fermi-Dirac distribution with zero neutrino mass and chemical potential is used to derive Eq. (\ref{eq:02b}). This assumption is reasonable as the neutrino mass is much smaller than ${\rm MeV}/c^2$ and the zero chemical potential is consistent with current observational constraints \cite{Pitrou2018.PhysRep.754.1}. The Fermi-Dirac distribution with $\mu_e+\mu_{e^+}=0$, where $\mu_{e^+}$ is the positron chemical potential, is used to derive the first equality of Eq. (\ref{eq:02c}). Here $\mu_e+\mu_{e^+}=0$ corresponds to the chemical equilibrium of the $e^-+e^+\leftrightarrow\gamma+\gamma$ reaction. The series expansion in the second equality of Eq. (\ref{eq:02c}) facilitates numerical calculations. The ideal gas model is used to derive Eq. (\ref{eq:02d}). The internal degeneracy of particles has been taken into account in the above results.

\begin{table*}[!t]
  \centering
  \caption{Nuclides considered in the nuclear network, together with mass excess $\Delta m_i$, mass number $A_i$, proton number $Z_i$ and spin $s_i$. The data comes from \cite{Audi2017.ChinPhysC.41.030001}. Mass excess is given in ${\rm MeV}$. Note that the nuclide mass $m_i=A_im_u+\Delta m_i-Z_im_e$, where $m_u$ is the atomic mass constant.}
  \label{tab:01}
  \begin{tabular*}{\hsize}{@{\ \ }@{\extracolsep{\fill}}llllllllll@{\ \ }}
    \toprule
    No. & 1 & 2 & 3 & 4 & 5 & 6 & 7 & 8 & 9 \\
    \hline
    Nuclide & $n$ & $p$ & $D$ & $T$ & ${^3{\rm He}}$ & ${^4{\rm He}}$ & ${^6{\rm Li}}$ & ${^7{\rm Li}}$ & ${^7{\rm Be}}$ \\
    $\Delta m_i$ & $8.0713$ & $7.2890$ & $13.1357$ & $14.9498$ & $14.9312$ & $2.4249$ & 14.0869 & $14.9071$ & $15.7690$ \\
    $A_i$ & 1 & 1 & 2 & 3 & 3 & 4 & 6 & 7 & 7 \\
    $Z_i$ & 0 & 1 & 1 & 1 & 2 & 2 & 3 & 3 & 4 \\
    $s_i$ & $1/2$ & $1/2$ & $1$ & $1/2$ & $1/2$ & $0$ & 1 & $3/2$ & $3/2$ \\
    \toprule
  \end{tabular*}
\end{table*}

\begin{table*}[!t]
  \centering
  \caption{The nuclear reaction network implemented in \texttt{BBNLab}. The $n\leftrightarrow p$ reaction includes three weak reactions (see Eq. (37) in \cite{Alpher1953.PhysRev.92.1347}).}
  \label{tab:02}
  \begin{tabular*}{\hsize}{@{\,}@{\extracolsep{\fill}}ll|ll|ll|ll@{\,}}
    \toprule
    No. & Reaction & No. & Reaction & No. & Reaction & No. & Reaction \\
    \hline
    01 & $n\leftrightarrow p$ & 09 & $T+D\leftrightarrow n+{^4{\rm He}}$ & 17 & ${^3{\rm He}}+{^3{\rm He}}\leftrightarrow p+p+{^4{\rm He}}$ & 25 & ${^6{\rm Li}}+D\leftrightarrow n+{^7{\rm Be}}$ \\
    02 & $p+n\leftrightarrow\gamma+D$ & 10 & $T+T\leftrightarrow n+n+{^4{\rm He}}$ & 18 & ${^4{\rm He}}+D\leftrightarrow\gamma+{^6{\rm Li}}$ & 26 & ${^6{\rm Li}}+D\leftrightarrow p+{^7{\rm Li}}$ \\
    03 & $D+n\leftrightarrow\gamma+T$ & 11 & ${^3{\rm He}}+n\leftrightarrow\gamma+{^4{\rm He}}$ & 19 & ${^4{\rm He}}+T\leftrightarrow\gamma+{^7{\rm Li}}$ & 27 & ${^7{\rm Li}}+p\leftrightarrow{^4{\rm He}}+{^4{\rm He}}$ \\
    04 & $D+p\leftrightarrow\gamma+{^3{\rm He}}$ & 12 & ${^3{\rm He}}+n\leftrightarrow p+T$ & 20 & ${^4{\rm He}}+{^3{\rm He}}\leftrightarrow\gamma+{^7{\rm Be}}$ & 28 & ${^7{\rm Li}}+p\leftrightarrow\gamma+{^4{\rm He}}+{^4{\rm He}}$ \\
    05 & $D+D\leftrightarrow n+{^3{\rm He}}$ & 13 & ${^3{\rm He}}+D\leftrightarrow p+{^4{\rm He}}$ & 21 & ${^6{\rm Li}}+n\leftrightarrow\gamma+{^7{\rm Li}}$ & 29 & ${^7{\rm Li}}+D\leftrightarrow n+{^4{\rm He}}+{^4{\rm He}}$ \\
    06 & $D+D\leftrightarrow p+T$ & 14 & ${^3{\rm He}}+T\leftrightarrow\gamma+{^6{\rm Li}}$ & 22 & ${^6{\rm Li}}+n\leftrightarrow T+{^4{\rm He}}$ & 30 & ${^7{\rm Be}}+n\leftrightarrow p+{^7{\rm Li}}$ \\
    07 & $T\rightarrow\bar{\nu}_e+e^-+{^3{\rm He}}$ & 15 & ${^3{\rm He}}+T\leftrightarrow D+{^4{\rm He}}$ & 23 & ${^6{\rm Li}}+p\leftrightarrow\gamma+{^7{\rm Be}}$ & 31 & ${^7{\rm Be}}+n\leftrightarrow{^4{\rm He}}+{^4{\rm He}}$ \\
    08 & $T+p\leftrightarrow\gamma+{^4{\rm He}}$ & 16 & ${^3{\rm He}}+T\leftrightarrow n+p+{^4{\rm He}}$ & 24 & ${^6{\rm Li}}+p\leftrightarrow{^3{\rm He}}+{^4{\rm He}}$ & 32 & ${^7{\rm Be}}+D\leftrightarrow p+{^4{\rm He}}+{^4{\rm He}}$ \\
    \toprule
  \end{tabular*}
\end{table*}

As we discussed before, aside from the gravitational interaction, the thermodynamics of the plasma and neutrinos, and the dynamics of the scalar field are independent of each other. Therefore, energy conservation provides three other evolution equations. For plasma, we obtain
\begin{equation}\label{eq:03}
  \dot{\rho}_{\rm plasma}+3H\left(\rho_{\rm plasma}+p_{\rm plasma}/c^2\right)=0,
\end{equation}
where $\rho_{\rm plasma}=\rho_\gamma+\rho_{e}+\rho_b$, $p_{\rm plasma}=p_\gamma+p_e+p_b$, $p_e=p_{e^-}+p_{e^+}$, and the pressures are given by
\begin{subequations}
\begin{align}
  p_\gamma &= \frac{1}{3}\rho_\gamma c^2, \\
  p_e      &= \frac{k_{\rm B}^4T_\gamma^4}{3\pi^2\hbar^3c^3}\int_{z_\gamma}^{\infty}(x^2-z_\gamma^2)^{3/2}\times \nonumber\\
           &  \qquad\qquad\qquad\left(\frac{1}{e^{x-\phi_e}+1}+\frac{1}{e^{x+\phi_e}+1}\right)\dx x, \nonumber\\
           &= \frac{2m_e^4c^5}{\pi^2\hbar^3z_\gamma}\sum_{i=1}^{\infty}\frac{(-1)^{i+1}}{i}\cosh(i\phi_e)L(iz_\gamma), \\
  p_b      &= n_bk_{\rm B}T_\gamma\sum_{i=1}^{N_{\rm nuclei}}Y_i,
\end{align}
\end{subequations}
where $L(x)$ is a special function defined in Appendix \ref{App:A}. The basic theory and assumptions used to derive the above results are discussed in the previous paragraph. For neutrinos, considering its pressure $p_\nu=\rho_\nu c^2/3$, we obtain
\begin{equation}\label{eq:05}
  \dot{T}_\nu+HT_\nu=0.
\end{equation}
Integrating the above equation gives $T_\nu\propto a^{-1}$. For the scalar field, the equation of energy conservation is exactly its equation of motion \cite{Tian2020.PRD.101.063531}
\begin{equation}\label{eq:06}
  \ddot{\phi}+3H\dot{\phi}+c^2V'=0,
\end{equation}
where $V'\equiv\dx V/\dx\phi$.

There are two other equations that are closely related to baryons. One is the equation of baryon number conservation
\begin{equation}\label{eq:07}
  \dot{n}_b+3Hn_b=0.
\end{equation}
The other is a constraint equation given by the electrical neutrality of the Universe, which can be written as
\begin{equation}\label{eq:08}
  \frac{n_{e^-}-n_{e^+}}{n_\gamma}=\eta\sum_{i=1}^{N_{\rm nuclei}}Z_iY_i,
\end{equation}
where $\eta\equiv n_b/n_\gamma$, and $n_{e^-}$, $n_{e^+}$, $n_\gamma$ are the number densities of relevant ingredients, $n_\gamma=2\zeta(3)m_e^3c^3/(\pi^2\hbar^3z_\gamma^3)$ and
\begin{align}\label{eq:09}
  \frac{n_{e^-}-n_{e^+}}{n_\gamma}
  &=\frac{\sinh\phi_e}{2\zeta(3)}\int_{z_\gamma}^{\infty}\frac{x(x^2-z_\gamma^2)^{1/2}}{\cosh x+\cosh\phi_e}\dx x \nonumber\\
  &=\frac{z_\gamma^3}{\zeta(3)}\sum_{i=1}^{\infty}(-1)^{i+1}\sinh(i\phi_e)L(iz_\gamma),
\end{align}
where $\zeta(x)$ is the Riemann zeta function and $\zeta(3)\approx1.2021$. So far, roughly speaking, the above results determine the thermodynamics of the Universe: There are six equations $\{$Eqs. (\ref{eq:01}), (\ref{eq:03}) and (\ref{eq:05}--\ref{eq:08})$\}$ and six variables $\{a,T_\gamma,T_\nu,\phi_e,n_b,\phi\}$. Note that it is unnecessary to find the solution of $a(t)$ because other equations depend directly on $H$ rather than $a$.

The rest is the nuclear reaction network which rules the evolution of $Y_i$. The general form of a nuclear reaction can be written as
\begin{widetext}
\begin{equation}\label{eq:10}
  N_{i_1}(^{A_{i_1}}Z_{i_1})+N_{i_2}(^{A_{i_2}}Z_{i_2})+\cdots+N_{i_p}(^{A_{i_p}}Z_{i_p})
  \longleftrightarrow N_{j_1}(^{A_{j_1}}Z_{j_1})+N_{j_2}(^{A_{j_2}}Z_{j_2})+\cdots+N_{j_q}(^{A_{j_q}}Z_{j_q}),
\end{equation}
where $^{A_{i}}Z_{i}$ denotes nuclide $i$ (see Table \ref{tab:01}), $N_i$ is the stoichiometric coefficient. The evolution of $Y_i$ takes the form \cite{Pitrou2018.PhysRep.754.1}
\begin{equation}\label{eq:11}
  \dot{Y}_{i_1}=\sum_{\textrm{reactions with } i_1}N_{i_1}\left(
  \Gamma_{j_1\cdots j_q\rightarrow i_1\cdots i_p}\frac{Y_{j_1}^{N_{j_1}}\cdots Y_{j_q}^{N_{j_q}}}{N_{j_1}!\cdots N_{j_q}!}
  -\Gamma_{i_1\cdots i_p\rightarrow j_1\cdots j_q}\frac{Y_{i_1}^{N_{i_1}}\cdots Y_{i_p}^{N_{i_p}}}{N_{i_1}!\cdots N_{i_p}!}\right)
  \equiv\Gamma_{i_1},
\end{equation}
where the sum includes all reactions involving nuclide $i_1$, $\Gamma_{i_1\cdots i_p\rightarrow j_1\cdots j_q}$ and $\Gamma_{j_1\cdots j_q\rightarrow i_1\cdots i_p}$ are the reaction rates (see Sec. \ref{sec:0201b}), and the last equality defines $\Gamma_i$. The nuclear reactions considered in this work are listed in Table \ref{tab:02}. These reactions are selected from Table 2 in \cite{Pisanti2008.ComputPhysCommun.178.956} with the criterion that the reactions involve only the nuclides listed in Table \ref{tab:01}. As we will see, this nuclear reaction network gives reasonable BBN predictions.
\end{widetext}

The above results form complete BBN evolution equations. However, Eqs. (\ref{eq:03}) and (\ref{eq:08}) are implicit and not suitable for direct numerical calculations. Here we convert these two equations into explicit form. Multiplying Eq. (\ref{eq:08}) by $\zeta(3)/z_\gamma^3$, differentiating the result with respect to $t$, and substituting Eq. (\ref{eq:11}) into the result to eliminate $\dot{Y}_i$, we obtain
\begin{equation}\label{eq:12}
  \kappa_1\dot{\phi}_e+\kappa_2\dot{z}_\gamma=\kappa_3,
\end{equation}
where the coefficients are given by
\begin{subequations}\label{eq:13}
\begin{align}
  \kappa_1 &= \sum_{i=1}^{\infty}(-1)^{i+1}i\cosh(i\phi_e)L(iz_\gamma), \\
  \kappa_2 &= \sum_{i=1}^{\infty}(-1)^{i+1}i\sinh(i\phi_e)L'(iz_\gamma), \\
  \kappa_3 &= \frac{\pi^2\hbar^3n_b}{2m_e^3c^3}\sum_{i=1}^{N_{\rm nuclei}}Z_i(\Gamma_i-3HY_i),
\end{align}
\end{subequations}
where $L'(x)\equiv\dx L/\dx x$ and its explicit expression is given in Appendix \ref{App:A}. Expanding $\dot{\rho}_{\rm plasma}$ by $\{\dot{\phi}_e,\dot{z}_\gamma,\dot{n}_b,\dot{Y}_i\}$, and substituting Eqs. (\ref{eq:07}) and (\ref{eq:11}) to eliminate $\{\dot{n}_b,\dot{Y}_i\}$, then Eq. (\ref{eq:03}) can be rewritten as
\begin{equation}\label{eq:14}
  \kappa_4\dot{\phi}_e+\kappa_5\dot{z}_\gamma=\kappa_6,
\end{equation}
where the coefficients are given by
\begin{subequations}\label{eq:15}
\begin{align}
  \kappa_4&=\sum_{i=1}^{\infty}(-1)^{i+1}i\sinh(i\phi_e)M(iz_\gamma), \\
  \kappa_5&=\frac{\pi^2\hbar^3}{2m_e^4c^3}\left(-\frac{4\rho_\gamma}{z_\gamma}-\frac{3m_en_b}{2z_\gamma^2}\sum_{i=1}^{N_{\rm nuclei}}Y_i\right) \nonumber\\
          & \qquad+\sum_{i=1}^{\infty}(-1)^{i+1}i\cosh(i\phi_e)M'(iz_\gamma), \\
  \kappa_6&=\frac{\pi^2\hbar^3}{2m_e^4c^3}\left[-3H\left(\rho_\gamma+\rho_e+\frac{p_{\rm plasma}}{c^2}\right)\right. \nonumber\\
          & \qquad\left.-n_b\sum_{i=1}^{N_{\rm nuclei}}\Gamma_i\left(m_i+\frac{3m_e}{2z_\gamma}\right)\right],
\end{align}
\end{subequations}
where $M'(x)\equiv\dx M/\dx x$ and its explicit expression is given in Appendix \ref{App:A}. The combination of Eqs. (\ref{eq:12}) and (\ref{eq:14}) gives
\begin{align}
  \dot{\phi}_e   &= \frac{\kappa_3\kappa_5-\kappa_2\kappa_6}{\kappa_1\kappa_5-\kappa_2\kappa_4}, \\
  \dot{z}_\gamma &= \frac{\kappa_1\kappa_6-\kappa_3\kappa_4}{\kappa_1\kappa_5-\kappa_2\kappa_4}.
\end{align}
So far, we can obtain the explicit expressions of the first time-derivative of $\{z_\gamma,z_\nu,\phi_e,\eta,\phi,\dot{\phi},Y_i\}$, where $z_\nu\equiv m_ec^2/(k_{\rm B}T_\nu)$. Here we replace $\{T_\nu,n_b\}$ with $\{z_\nu,\eta\}$ to make the variables dimensionless. In addition, $\dot{z}_\nu=Hz_\nu$ and $\dot{\eta}=3\pi^2\hbar^3z_\gamma^2n_b(\dot{z}_\gamma-Hz_\gamma)/[2\zeta(3)m_e^3c^3]$  can be easily derived from the previous equations.

\texttt{BBNLab} adopts the variables $\{z_\gamma,z_\nu,\phi_e,\eta,\phi,\dot{\phi},Y_i\}$ in its numerical calculations. Furthermore, we change the ``time" variable from $t$ to the $e$-folding number $N$, where $N\equiv\ln(a/a_1)$, $a_1$ is the scale factor at one specific time and its absolute value is meaningless. The transformation of the evolution equations directly follows $\dx X/\dx N=\dot{X}/H$. Hereafter we use the subscript ``ini" to denote the beginning of BBN. If we set the initial temperature to $10^{12}\,{\rm K}$, then $N\in[N_{\rm ini},N_{\rm ini}+10]$ roughly corresponds to $10^{12}\,{\rm K}\leq T_\gamma\lesssim10^{8}\,{\rm K}$. The exact value of $N_{\rm ini}$ have no physical effect as the BBN evolution equations do not depend on $N$ explicitly. The reason for keeping $N_{\rm ini}$ here is that it is an auxiliary value used to calculate the initial conditions of the stepwise scalar field (see Sec. \ref{sec:0202}). For clarity, we set $N_{\rm ini}=0$ for the standard model and the exponential scalar field model (see Sec. \ref{sec:0201a}). The BBN evolution equations are stiff. In \texttt{BBNLab}, we use the \texttt{Matlab/ode23s} solver with suitable option structure to integrate this stiff system. We set the options as \texttt{odeset( 'RelTol',1E-10,'AbsTol',1E-10,'MaxStep',1E-3)} for Fig. \ref{fig:01}, and \texttt{odeset('RelTol',1E-10,'AbsTol', 1E-10,'MaxStep',2E-4)} for Fig. \ref{fig:03} and Fig. \ref{fig:04}. Details about \texttt{ode23s} can be found in \url{https://www.mathworks.com/help/matlab/ref/ode23s.html}. For all series expansions involved in our numerical calculations, e.g., Eq. (\ref{eq:02c}), we adopt the results up to the 20th order.

\subsubsection{Models of the scalar field}\label{sec:0201a}
The goal of this paper is to study BBN in the framework of the stepwise scalar field model proposed in \cite{Tian2020.PRD.101.063531}. As a comparison, we will also present the BBN results for the standard model, in which no scalar fields exist, and the exponential scalar field model \cite{Ratra1988.PRD.37.3406,Copeland1998.PRD.57.4686}. Note that the stepwise scalar field model is inspired by the exponential scalar field model and these two models share some similar features, e.g., the scaling property \cite{Tian2020.PRD.102.063509}. Discussions on the exponential scalar field model may shed light on what kind of the stepwise scalar field model can survive from the BBN constraints.

In our conventions \cite{Tian2020.PRD.101.063531}, the exponential potential reads
\begin{equation}
  V(\phi)=V_0\exp(-\lambda\phi),
\end{equation}
where $\lambda$ is a dimensionless constant, $V_0$ is a constant in unit of ${\rm length}^{-2}$, $\phi$ is dimensionless. Scaling solution is a main mathematical property of this model \cite{Copeland1998.PRD.57.4686}. In the radiation-dominated Universe, the scaling solution requires $\lambda>2$. Furthermore, previous discussions on BBN require $\lambda\gtrsim9$ \cite{Bean2001.PRD.64.103508}. The potential of the stepwise scalar field model is written as \cite{Tian2020.PRD.101.063531}
\begin{equation}
  V(\phi)=V_0\exp\left[-\frac{\lambda_1+\lambda_2}{2}\phi-\frac{\alpha(\lambda_1-\lambda_2)}{2}\sin\frac{\phi}{\alpha}\right],
\end{equation}
where $\lambda_1$, $\lambda_2$ and $\alpha$ are dimensionless constants, $V_0$ and $\phi$ follow the previous conventions. Period-doubling bifurcation together with the oscillating scaling and chaotic accelerating solutions are the main mathematical properties of this model \cite{Tian2020.PRD.102.063509}. In order to explain the cosmic late-time acceleration and solve the accompanying coincidence problem, we require $\{\lambda_1+\lambda_2>4,\,0<\lambda_2\lesssim0.4,\,\alpha=\mathcal{O}(1)\}$ \cite{Tian2020.PRD.101.063531}. Furthermore, we require $\lambda_2=\mathcal{O}(10^{-4})$ to unify inflation \cite{Tian2021.PRD.103.123545}.

\subsubsection{Nuclear reaction rates}\label{sec:0201b}
For clarity, we adopt the conventions that the forward direction of the reactions listed in Table \ref{tab:02} and Eq. (\ref{eq:10}) is taken as left to right. For the $n\leftrightarrow p$ reaction, weak interaction theory gives \cite{Alpher1953.PhysRev.92.1347,Beaudet1976.AstronAstrophys.49.415}
\begin{subequations}\label{eq:20}
\begin{align}
  \Gamma_{n\rightarrow p} &= K\int_{1}^{\infty}\frac{x(x+q)^2\sqrt{x^2-1}}{(1+e^{xz_\gamma})\left[1+e^{-(x+q)z_\nu}\right]}\dx x \nonumber\\
    &\quad + K\int_{1}^{\infty}\frac{x(x-q)^2\sqrt{x^2-1}}{(1+e^{-xz_\gamma})\left[1+e^{(x-q)z_\nu}\right]}\dx x, \\
  \Gamma_{p\rightarrow n} &= K\int_{1}^{\infty}\frac{x(x-q)^2\sqrt{x^2-1}}{(1+e^{xz_\gamma})\left[1+e^{-(x-q)z_\nu}\right]}\dx x \nonumber\\
    &\quad + K\int_{1}^{\infty}\frac{x(x+q)^2\sqrt{x^2-1}}{(1+e^{-xz_\gamma})\left[1+e^{(x+q)z_\nu}\right]}\dx x,
\end{align}
\end{subequations}
where $q=(m_n-m_p)/m_e\approx2.5310$, $m_n$ and $m_p$ are the neutron and proton mass respectively, $K=6.9503\times10^{-4}\,{\rm s}^{-1}$ is the normalization constant computed by requiring $\Gamma_{n\rightarrow p}=1/\tau_n$ at low temperatures, $\tau_n=879.4\,{\rm s}$ is the neutron lifetime \cite{Czarnecki2018.PRL.120.202002}. The Fermi correction discussed in \cite{Beaudet1976.AstronAstrophys.49.415} is ignored here. At high temperatures, we have $z_\nu=z_\gamma$ so that $\Gamma_{n\rightarrow p}=\Gamma_{p\rightarrow n}e^{qz_\gamma}$, which is consistent with the equilibrium abundances (see Eq. (\ref{eq:25})). For other reactions, generally, the forward thermonuclear reaction rates in the CGS units are given in the literature \cite{Pitrou2018.PhysRep.754.1}. The forward reaction rates read \cite{Pitrou2018.PhysRep.754.1}
\begin{equation}\label{eq:21}
  \Gamma_{i_1\cdots i_p\rightarrow j_1\cdots j_q}=\left[\frac{n_bm_u}{{\rm g}/{\rm cm}^3}\right]^{(N_{i_1}+\cdots+N_{i_p})-1}
          \cdot R_{\rm t}\cdot[{\rm s}^{-1}],
\end{equation}
where $R_{\rm t}$ is the numerical value of the thermonuclear reaction rate in CGS units. The $R_{\rm t}$ actually used in the present calculation and relevant references are presented in the code. Most of the reaction rates we use are consistent with those used in \texttt{PRIMAT} (Version 0.2.0) \cite{Pitrou2021.MNRAS.502.2474}. Especially, as did in \cite{Pitrou2021.MNRAS.502.2474}, we adopt the most recent measurements about the $D+p\rightarrow\gamma+{^3{\rm He}}$ reaction \cite{Mossa2020.Nature.587.210}. The deuterium abundance is sensitive to this reaction \cite{Pitrou2021.MNRAS.502.2474,Pitrou2021.NatRevPhys.3.231} and is the core in our discussions. The last term in Eq. (\ref{eq:21}) is the physical unit, i.e., the unit of $\Gamma_{i_1\cdots i_p\rightarrow j_1\cdots j_q}$. The inverse reaction rates read \cite{Pitrou2018.PhysRep.754.1}
\begin{align}
  \Gamma_{j_1\cdots j_q\rightarrow i_1\cdots i_p}
  &=\left[\frac{n_bm_u}{{\rm g}/{\rm cm}^3}\right]^{(N_{j_1}+\cdots+N_{j_q})-(N_{i_1}+\cdots+N_{i_p})}\times \nonumber\\
  &\qquad\frac{\gamma_{j_1\cdots j_q\rightarrow i_1\cdots i_p}}{\gamma_{i_1\cdots i_p\rightarrow j_1\cdots j_q}}
  \times\Gamma_{i_1\cdots i_p\rightarrow j_1\cdots j_q},
\end{align}
where the dimensionless ratio
\begin{equation}
  \frac{\gamma_{j_1\cdots j_q\rightarrow i_1\cdots i_p}}{\gamma_{i_1\cdots i_p\rightarrow j_1\cdots j_q}}
  =\beta_1\left(\frac{T_\gamma}{10^9\,{\rm K}}\right)^{\beta_2}\exp\left(\frac{\beta_3\cdot10^9\,{\rm K}}{T_\gamma}\right),
\end{equation}
and the constant $\beta_i$ is given in our code. Note that each nuclear reaction corresponds to a set of $\beta_i$. Uncertainties \cite{Krauss1990.ApJ.358.47} of the reaction rates are not included in the present calculation. More nuclear reactions and the uncertainties will be considered in the future.

\subsection{Initial conditions}\label{sec:0202}
Considering the neutrino-plasma thermal equilibrium, we set the initial temperatures $T_{\gamma,{\rm ini}}=T_{\nu,{\rm ini}}=10^{12}\,{\rm K}$ in \texttt{BBNLab}. For the baryon-to-photon number ratio, we set $\eta_{\rm ini}\in11/4\times[10^{-10},10^{-9}]$ so that $\eta\in[10^{-10},10^{-9}]$ at the ending of BBN. Here the coefficient $11/4$ is due to the entropy transfer from $e^\pm$ pairs to photons \cite{Pitrou2018.PhysRep.754.1}.

The initial high temperature makes all nuclides in statistical equilibrium, which includes thermal and chemical equilibrium. The initial number density of nuclide $i$ reads
\begin{align}\label{eq:24}
  n_{i,{\rm ini}}=\frac{g_i}{\hbar^3}\left(\frac{m_ik_{\rm B}T_{\gamma,{\rm ini}}}{2\pi}\right)^{3/2}
    \exp\left(\frac{\mu_{i,{\rm ini}}-m_ic^2}{k_{\rm B}T_{\gamma,{\rm ini}}}\right),
\end{align}
where $g_i=2s_i+1$ is the spin degeneracy, $\mu_{i,{\rm ini}}$ is the initial chemical potential. The chemical potential of leptons is negligible compared to that of baryons. Therefore we obtain $\mu_{n,{\rm ini}}=\mu_{p,{\rm ini}}$ based on the $n\leftrightarrow p$ reaction. Applying Eq. (\ref{eq:24}) to neutrons and protons, we obtain
\begin{equation}\label{eq:25}
  \frac{n_{n,{\rm ini}}}{n_{p,{\rm ini}}}=\left(\frac{m_n}{m_p}\right)^{3/2}\exp\left[\frac{(m_p-m_n)c^2}{k_{\rm B}T_{\gamma,{\rm ini}}}\right].
\end{equation}
Initially, neutrons and protons are much more numerous than other nucleons. Therefore, it is reasonable to adopt $n_{n,{\rm ini}}+n_{p,{\rm ini}}=n_{b,{\rm ini}}$, which in turn gives
\begin{subequations}\label{eq:26}
\begin{align}
  Y_{n,{\rm ini}}&=1/(1+n_{p,{\rm ini}}/n_{n,{\rm ini}}), \\
  Y_{p,{\rm ini}}&=1-Y_{n,{\rm ini}}.
\end{align}
\end{subequations}
For other nuclides, we set $Y_{i,{\rm ini}}=0$ in \texttt{BBNLab}. Note that we turn off all nuclear reactions except the $n\leftrightarrow p$ reaction until $T_\gamma=10^{10}\,{\rm K}$. Once the full nuclear reaction network is turned on, the heavier nuclides reach their equilibrium abundances very quickly (see Fig. \ref{fig:01} for an illustration). Therefore, it is reasonable to set $Y_{i,{\rm ini}}=0$ for the heavier nuclides and the final results is independent of this setting. This strategy is also adopted by \texttt{PRIMAT} \cite{Pitrou2018.PhysRep.754.1}.

The initial value of $\phi_e$ can be obtained by numerically solving Eq. (\ref{eq:08}). Here we present a high temperature approximate solution. The right side of Eq. (\ref{eq:08}) is much less than $1$ as $\eta\ll1$. For its left side, i.e., Eq. (\ref{eq:09}), one can verify that $\lim_{z_\gamma\rightarrow0}z_\gamma^3L(iz_\gamma)=2/i^3=\mathcal{O}(1)$, where $z_\gamma\rightarrow0$ corresponds to the high temperature approximation. Therefore, Eq. (\ref{eq:08}) indicates $\sinh(i\phi_e)=\mathcal{O}(\eta)\ll1$ at the beginning of BBN. Substituting the Taylor expansion $\sinh x=x$ into Eq. (\ref{eq:08}), we obtain
\begin{align}
  \phi_{e,{\rm ini}}=\frac{\zeta(3)\eta_{\rm ini}\sum_{i=1}^{N_{\rm nuclei}}Z_iY_{i,{\rm ini}}}{z_{\gamma,{\rm ini}}^3\sum_{i=1}^{\infty}(-1)^{i+1}i L(iz_{\gamma,{\rm ini}})}.
\end{align}
This result is consistent with Eq. (20) in \cite{Arbey2012.ComputPhysCommun.183.1822}.

The initial conditions of the scalar field, i.e., in principle, the values of $\{\phi_{\rm ini},\dot{\phi}_{\rm ini},V_0\}$, are calculated as follows. For the exponential scalar field model, we only consider the scaling case and assume that the scalar field reaches the scaling attractor at the beginning of BBN. Without loss of generality, we can set $\phi_{\rm ini}=0$ and use $V_0$ to adjust the potential energy. Then the initial conditions can be calculated as follows
\begin{enumerate}
  \item $x_{1,{\rm ini}}=4/(\sqrt{6}\lambda)$, \\
    $x_{2,{\rm ini}}=2/(\sqrt{3}\lambda)$,
  \item $\Omega_{\phi,{\rm ini}}=4/\lambda^2$, \\
    $\rho_{\rm matter,ini}=\rho_{\rm plasma,ini}+\rho_{\nu,{\rm ini}}$, \\
    $\rho_{\rm total,ini}=\rho_{\rm matter,ini}/(1-\Omega_{\phi,{\rm ini}})$, \\
    $H_{\rm ini}=(8\pi G\rho_{\rm total,ini}/3)^{1/2}$,
  \item $\dot{\phi}_{\rm ini}=\sqrt{6}H_{\rm ini}x_{1,{\rm ini}}$, \\
    $V_0=3(H_{\rm ini}x_{2,{\rm ini}}/c)^2$.
\end{enumerate}
Here $x_1$ and $x_2$ are the dimensionless variables defined in the dynamical system analysis (see Eq. (7a) in \cite{Tian2020.PRD.101.063531}) and their initial values are given by the scaling solution, $\Omega_\phi$ is the relative energy density of the scalar field. For the stepwise scalar field model, instead of the scaling solution, we can use the numerical solution of the dynamical system to give the values of $x_1$ and $x_2$ at the beginning of BBN. The dynamical system is given by Eq. (8) with $w_{\rm m}=1/3$ in \cite{Tian2020.PRD.101.063531} and we do not repeat the equations and conventions here. Other subsequent steps are similar to the exponential case. However, one difference is that the value of $\phi_{\rm ini}$ is no longer fixed. Without loss of generality, we assume $0\leq\phi_{\rm ini}\leq2\alpha\pi$ and use $V_0$ to adjust the potential energy. Details are as follows
\begin{enumerate}
  \item Integrating the dynamical system in $N\in[0,N_{\rm ini}]$ gives the values of $\{x_{1,{\rm ini}},\,x_{2,{\rm ini}},\,\lambda_{\rm ini},\,\nu_{\rm ini}\}$, i.e., the values of $\{x_1,x_2,\lambda,\nu\}$ at $N=N_{\rm ini}$ (see Fig. \ref{fig:02} for an illustration). In this step, we set $\{x_1=0.75,x_2=0.5,\lambda=\lambda_2+2,\nu=\nu_+(\lambda)\}$ at $N=0$ for all following calculations. Compared to setting the values at $N=N_{\rm ini}$ directly, this step allows the system to enter the possible attractor solution and makes the desired values more natural. Note that $\lambda$ is an variable in the stepwise scalar field model, while it is a constant in the exponential scalar field model; $N_{\rm ini}$ is an auxiliary parameter and its absolute value have no physical meaning (see also discussions in Sec. \ref{sec:0201}).
  \item $\Omega_{\phi,{\rm ini}}=x_{1,{\rm ini}}^2+x_{2,{\rm ini}}^2$, \\
    $\rho_{\rm matter,ini}=\rho_{\rm plasma,ini}+\rho_{\nu,{\rm ini}}$, \\
    $\rho_{\rm total,ini}=\rho_{\rm matter,ini}/(1-\Omega_{\phi,{\rm ini}})$, \\
    $H_{\rm ini}=(8\pi G\rho_{\rm total,ini}/3)^{1/2}$,
  \item $\phi_{\rm ini}=\alpha\arccos\displaystyle\left[\frac{2}{\lambda_1-\lambda_2}\cdot\left(\lambda_{\rm ini}-\frac{\lambda_1+\lambda_2}{2}\right)\right]$, \\
    \texttt{if} $\nu_{\rm ini}>0$ \texttt{then} $\phi_{\rm ini}=2\alpha\pi-\phi_{\rm ini}$,
  \item $\dot{\phi}_{\rm ini}=\sqrt{6}H_{\rm ini}x_{1,{\rm ini}}$, \\
    $V_{\rm ini}=3(H_{\rm ini}x_{2,{\rm ini}}/c)^2$, \\
    $V_0=V_{\rm ini}\exp\displaystyle\left[\frac{\lambda_1+\lambda_2}{2}\phi_{\rm ini}+\frac{\alpha(\lambda_1-\lambda_2)}{2}\sin\frac{\phi_{\rm ini}}{\alpha}\right]$.
\end{enumerate}
Here the third step gives the solution of Eqs. (7b) and (7c) in \cite{Tian2020.PRD.101.063531}. Note that the function \texttt{arccos} returns values in the interval $[0,\pi]$ and thus the final result satisfies our initial assumption $0\leq\phi_{\rm ini}\leq2\alpha\pi$. Considering the initial temperature, the above procedure gives $V_0\sim G(100\,{\rm MeV})^4/\hbar^3c^7\sim10^{-80}l_{\rm P}^{-2}$, where $l_{\rm P}$ is the Planck length. This result is much smaller than that given in \cite{Tian2021.PRD.103.123545}. However, there is no inconsistency in the model. The reason is that there is a degeneracy between the absolute values of $V_0$ and $\phi_{\rm ini}$. Because the function $\sin$ is periodic, increasing $\phi_{\rm ini}$ by several periods together with a corresponding rescaled $V_0$ does not affect the potential and dynamics of the scalar field. In \cite{Tian2021.PRD.103.123545}, we assumed $\phi\approx\alpha\pi$ at the beginning of inflation and obtained $V_0\sim10^{-14}l_{\rm P}^{-2}$. If we want to restore these results, what we need to do is increase the above obtained $\phi_{\rm ini}$ by several periods so that $\exp[-(\lambda_1+\lambda_2)\phi_{\rm ini}/2]\sim10^{-66}$. The BBN results presented in this paper are independent of these settings.

\section{Numerical results}\label{sec:03}
To compare the theoretical predictions and observations, we consider $Y_{\rm P}\equiv4\,Y_{^4{\rm He}}$ (the ratio of the abundance by mass of helium-4), ${\rm D}/{\rm H}$, ${^3{\rm He}}/{\rm H}$ and ${^7{\rm Li}}/{\rm H}$ (the ratio of the abundance of the isotope to hydrogen, and $i/{\rm H}\equiv Y_i/Y_p$). Observational results are $Y_{\rm P}=0.2453\pm0.0034$ \cite{Aver2021.JCAP.03.027}, ${\rm D}/{\rm H}=(2.527\pm0.030)\times10^{-5}$ \cite{Cooke2018.ApJ.855.102}, ${^3{\rm He}}/{\rm H}<(1.1\pm0.2)\times10^{-5}$ \cite{Bania2002.Nature.415.54}, and ${^7{\rm Li}}/{\rm H}=(1.58_{-0.28}^{+0.35})\times10^{-10}$ \cite{Sbordone2010.AstronAstrophys.522.A26}. The ${^3{\rm He}}$ result is given by its abundance in our Galaxy and the result is adopted as an upper limit of the primordial abundance because the post-BBN evolution of ${^3{\rm He}}$ is unclear \cite{Bania2002.Nature.415.54,Pitrou2018.PhysRep.754.1}. As stated in Sec. \ref{sec:0202}, we compute the abundances for a given scalar field model over the range $\eta_{\rm today}\in[10^{-10},10^{-9}]$, which covers the \textit{Planck} 2018 result $\eta_{\rm today}=(6.1374\pm0.0383)\times10^{-10}$ \cite{Aghanim2020.AstronAstrophys.641.A6,Pitrou2021.MNRAS.502.2474}. Here the subscript ``today" means the present epoch or equivalently the ending of BBN.

In order to check the correctness of our code, we present the equilibrium abundances. Omitting the subscript ``ini" in Eq. (\ref{eq:26}) gives the equilibrium abundances of neutron and proton. This result is available at $T_\gamma\sim10^{12}\,{\rm K}$. For other nuclides, the equilibrium abundances are given by \cite{Pitrou2018.PhysRep.754.1}
\begin{align}
  & Y_{i,{\rm eq}} = g_i2^{(3A_i-5)/2}\pi^{(1-A_i)/2}\zeta(3)^{A_i-1}c^{3(1-A_i)}\eta^{A_i-1}\times\nonumber\\
  & \quad Y_p^{Z_i}Y_n^{A_i-Z_i}\left[\frac{m_i(k_{\rm B}T_\gamma)^{A_i-1}}{m_p^{Z_i}m_n^{A_i-Z_i}}\right]^{3/2}
  \exp\left(\frac{B_i}{k_{\rm B}T_\gamma}\right),
\end{align}
where the binding energy $B_i\equiv [Z_im_p+(A_i-Z_i)m_n-m_i]\times c^2$, and the relevant values are given in Table \ref{tab:01}. If the temperature is low enough that the neutrons and protons are not in thermal equilibrium, e.g., $T_\gamma\sim10^{10}\,{\rm K}$, then $Y_{i,{\rm eq}}$ should be calculated based on the exact values of $Y_n$ and $Y_p$, rather than their equilibrium values.

\subsection{Exponential scalar field}\label{sec:0301}
\begin{figure*}[!t]
  \centering
  \includegraphics[width=0.49\linewidth]{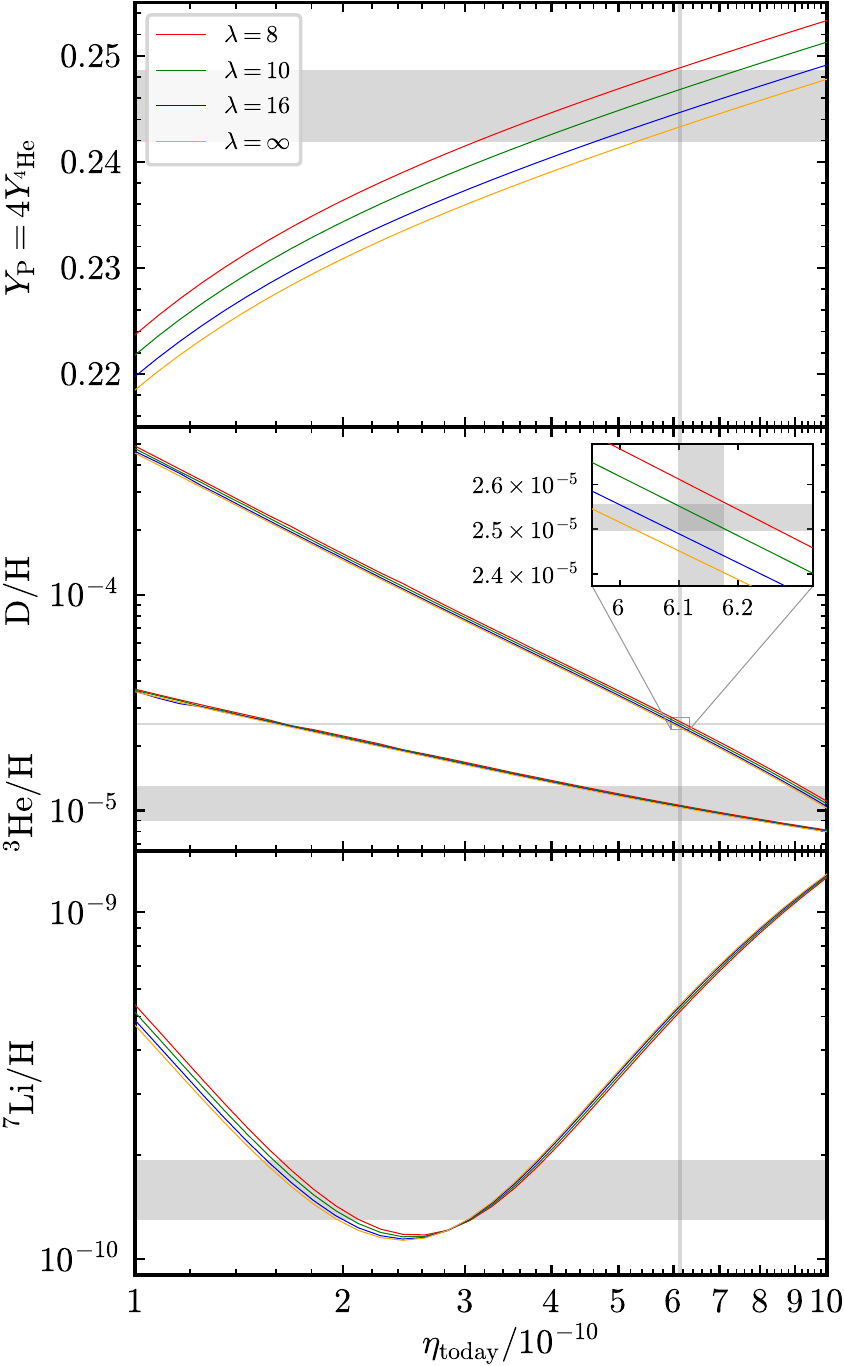}
  \includegraphics[width=0.49\linewidth]{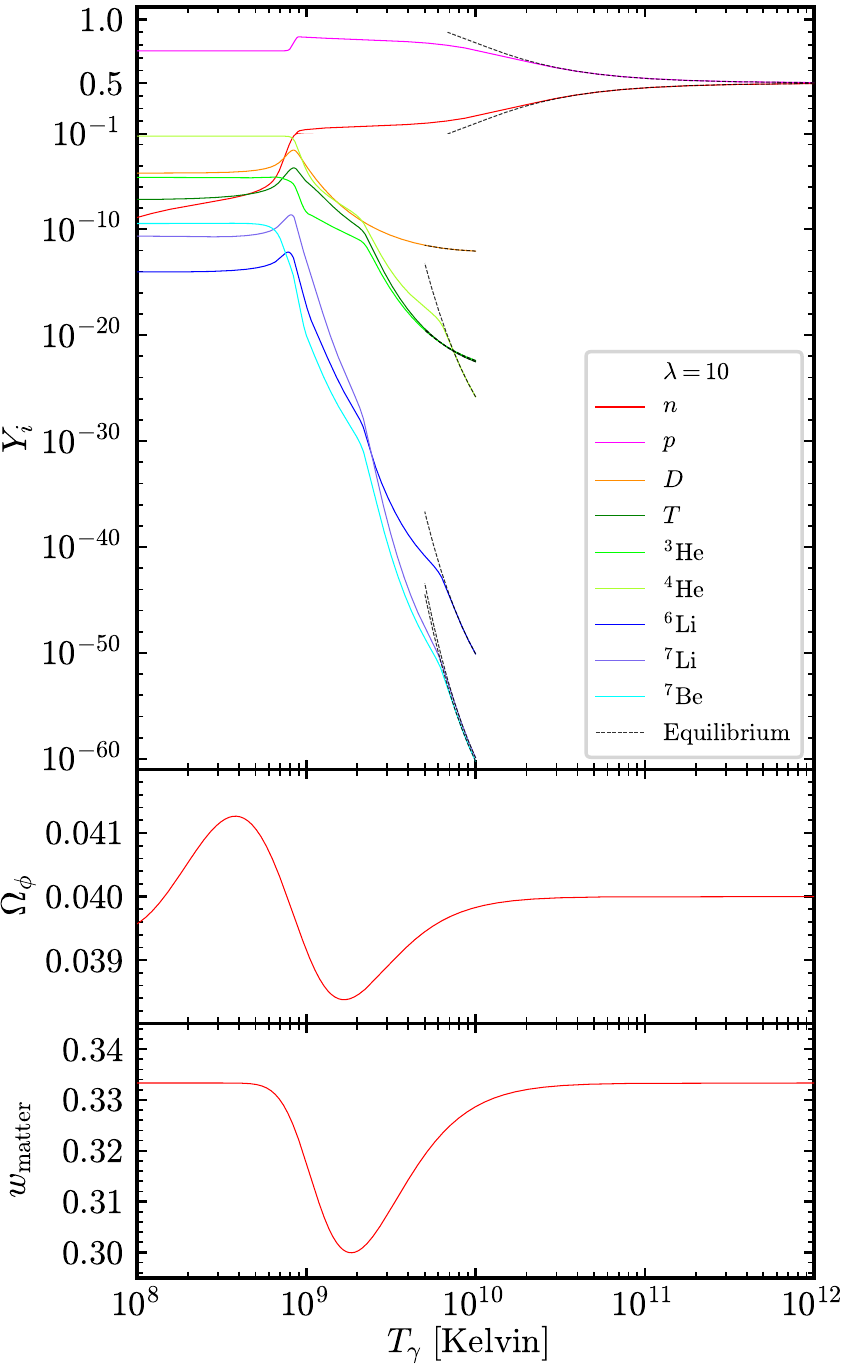}
  \caption{Left: Dependence of $\{Y_{\rm P},{\rm D}/{\rm H},{^3{\rm He}}/{\rm H},{^7{\rm Li}}/{\rm H}\}$ in $\eta_{\rm today}$ for the exponential scalar field model ($\lambda=8,10,16,\infty$) and observational constraints. Due to radioactive decay, the ${\rm T}$ and ${^7{\rm Be}}$ abundance has been added into ${^3{\rm He}}$ and ${^7{\rm Li}}$, respectively. Horizontal shaded areas represent the observational abundances, while vertical shaded areas represent the \textit{Planck} 2018 result on $\eta_{\rm today}$ (see the main text for the values). Right: Evolution of the elements abundances, $\Omega_\phi$ and $w_{\rm matter}$ for $\lambda=10$ and $\eta_{\rm today}=6.1374\times10^{-10}$. Dashed curves represent the equilibrium abundances. The fluctuations in $\Omega_\phi$ and $w_{\rm matter}$ evolutions are caused by the $e^\pm$ annihilation.}
  \label{fig:01}
\end{figure*}

Figure \ref{fig:01} presents the main BBN results for the exponential scalar field model. The predicted abundances as a function of $\eta_{\rm today}$ and the observed results are plotted in the left part. The case of $\lambda=\infty$ corresponds to the standard BBN model, at which $\Omega_\phi=0$. Compared to \texttt{PRIMAT}, our code gives a slightly lower ${^4{\rm He}}$ abundance, and consistent abundances of ${\rm D}$, ${^3{\rm He}}$ and ${^7{\rm Li}}$. The lower $Y_{\rm P}$ should be due to the incomplete nuclear reaction network. However, the difference is small relative to the observational uncertainty and thus is ignorable. Importantly, our code recovers the deuterium problem, which is first pointed out in \cite{Pitrou2021.MNRAS.502.2474} and states that the predicted ${\rm D}/{\rm H}$ abundance is lower than the observational constraints. Using a scalar field to solve the deuterium problem is the core in our following discussions. Decreasing $\lambda$, that is, increasing $\Omega_\phi$ during BBN, can significantly increase the abundances of ${^4{\rm He}}$ and ${\rm D}$. In particular, we can solve the deuterium problem with $\lambda=10$. A potential disadvantage here is that the predicted abundance of ${^4{\rm He}}$ may exceed its observed value if we consider the full \texttt{PRIMAT} nuclear reaction network. The case of $\lambda=8$ gives too large $Y_{\rm P}$ and ${\rm D}/{\rm H}$ and will be ruled out by the observations. This is consistent with the result given by \cite{Bean2001.PRD.64.103508} ($\lambda>9$ at $2\sigma$ C.L.). The abundances of ${^3{\rm He}}$ and ${^7{\rm Li}}$ remain almost unchanged when $\lambda$ changes from infinity to $8$. Consistent with conventional cognition \cite{Arbey2019.JCAP.11.038}, our results show that the exponential scalar field cannot solve the primordial lithium problem \cite{Fields2011.ARNPS.61.47}.

The case of $\lambda=10$ deserves more discussion as it can solve the deuterium problem. The evolution of the isotopes as a function of $T_\gamma$ and their equilibrium abundance are depicted in the right part of Fig. \ref{fig:01}. As the Universe expands (temperature decreases), the evolution of neutron and proton follow their equilibrium abundance until $T_\gamma\approx2\times10^{10}\,{\rm K}$. Heavier nuclei are taken into account when $T_\gamma=10^{10}\,{\rm K}$ and their evolution follow the equilibrium abundance until $T_\gamma\approx6\times10^{9}\,{\rm K}$. Therefore, it is reasonable for \texttt{BBNLab} to consider heavier nuclei starting from $T_\gamma=10^{10}\,{\rm K}$. We also plot the evolution of $\Omega_\phi$ and the equation of state (EOS) $w_{\rm matter}$. During BBN era, $\Omega_\phi$ remains almost constant ($4\%$), which corresponds to the scaling solution \cite{Copeland1998.PRD.57.4686}. Small fluctuations in $\Omega_\phi$ are due to the fluctuations in $w_{\rm matter}$, which in turn results from the $e^\pm$ annihilation. Combining the left and right parts of Fig. \ref{fig:01}, we conclude that if the relative scalar field energy density reaches $4\%$, then it can result in observable influences on the final BBN predictions. During BBN, such a scalar field may can affect the out-of-equilibrium process of neutron and proton around $T_\gamma\approx2\times10^{10}\,{\rm K}$, and the heavier nuclei evolution when $T_\gamma\lesssim6\times10^{9}\,{\rm K}$.

\subsection{Stepwise scalar field}\label{sec:0302}
In the stepwise scalar field model, our previous work established the following unified cosmic evolution scenario: inflation \cite{Tian2021.PRD.103.123545}, deflationary phase with very stiff EOS \cite{Tian2021.PRD.103.123545}, radiation era with oscillating scaling solution (OSS) \cite{Tian2020.PRD.102.063509}, and matter era with chaotic accelerating solution (CAS) \cite{Tian2020.PRD.101.063531,Tian2020.PRD.102.063509}. OSS during radiation era can attenuate the sensitive dependence of the late-time cosmic evolution on the early initial conditions and facilitate the cosmological parameter constraints \cite{Tian2020.PRD.102.063509}. This is a pragmatic choice rather than an observational confirmation. In principle, the Universe can also evolve as a CAS during the radiation era. The BBN discussions may shed light on the real type of cosmic evolution during the radiation era.

\begin{figure*}[!t]
  \centering
  \includegraphics[width=0.99\linewidth]{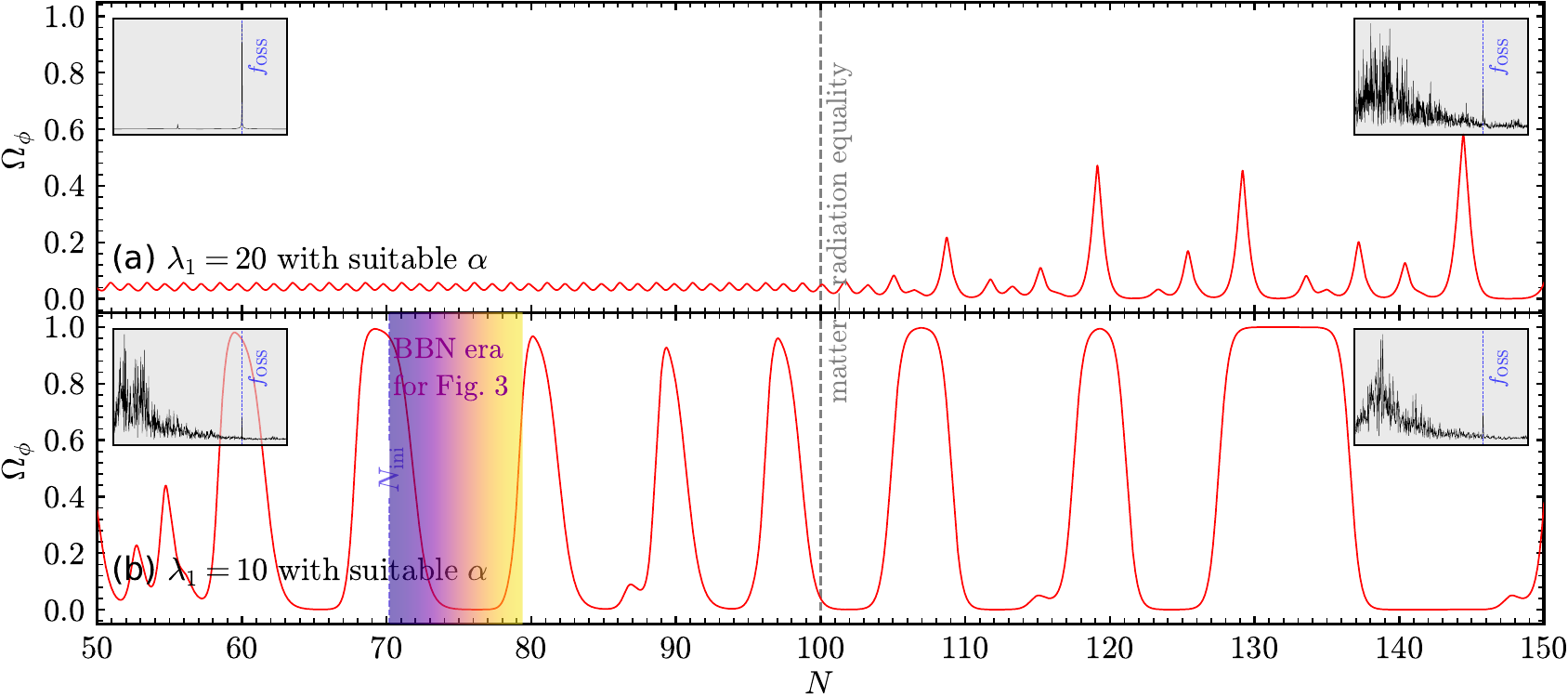}
  \caption{Evolution of the dark energy relative density $\Omega_\phi$ for the stepwise scalar field model. The calculations are based on Eq. (8) with $w_{\rm m}(N)=(1/3)/[1+\exp(N-N_{\rm eq})]$ in \cite{Tian2020.PRD.101.063531}. This equation describes the evolution of the Universe from radiation era to matter era \cite{Tian2020.PRD.102.063509}. $N=N_{\rm eq}$ corresponds to matter-radiation equality and we set $N_{\rm eq}=100$. We perform numerical calculations in $N\in[0,150]$ and plot the results within $N\in[50,150]$. The model parameters are $\lambda_1=20,\lambda_2=0.01$ and $\alpha=0.08$ for the subplot (a), and $\lambda_1=10$, $\lambda_2=0.01$, $\alpha=0.24$ for the subplot (b). To increase the robustness of our numerical algorithm, we set $\lambda_2=0.01$, rather than $10^{-4}$, which is favored by discussions on inflation \cite{Tian2021.PRD.103.123545}. The initial conditions are $x_{1,0}=0.75$, $x_{2,0}=0.5$, $\lambda_0=\lambda_2+2$ and $\nu_0=\nu_+(\lambda_0)$, which are the same as the settings in Sec. \ref{sec:0202}. The colored shaded region illustrates how to set the initial conditions of the stepwise scalar field at $T_\gamma=10^{12}\,{\rm K}$ in \texttt{BBNLab} (see discussions in the main text). The insets are the Fourier transform of $\Omega_\phi(N)$ with constant $w_{\rm m}$ ($w_{\rm m}=1/3$ for the left insets and $w_{\rm m}=0$ for the right insets). To plot this, we numerically solve the evolution equation in $N\in[0,1000]$ and perform Fourier transform in $N\in[100,1000]$. The initial conditions are the same as before. The blue vertical dashed lines denote $f_\textsc{oss}=3(1+w_{\rm m})/[\alpha\pi(\lambda_1+\lambda_2)]$ \cite{Tian2020.PRD.102.063509}. In the Fourier plots, regular peaks correspond to the OSS, while random forest corresponds to the CAS.}
  \label{fig:02}
\end{figure*}

We start from the OSS. Roughly speaking, Sec. \ref{sec:0301} shows that $\Omega_\phi\approx4\%$ is a critical point for the BBN constraints. Higher $\Omega_\phi$ would be excluded from observations, while lower $\Omega_\phi$ would have no observable effect. If we expect OSS in the radiation era, then we require $\lambda_1\gtrsim20$ because $\Omega_\phi\approx16/(\lambda_1+\lambda_2)^2$ \cite{Tian2020.PRD.101.063531} and $\lambda_2\ll1$ \cite{Tian2021.PRD.103.123545}. In particular, $\lambda_1\approx20$ may can be used to solve the deuterium problem as the exponential scalar field did. In Fig. \ref{fig:02} (a), we plot the evolution of $\Omega_\phi(N)$ in the radiation and matter epochs for $\lambda_1=20$. Other parameter settings and calculation details can be found in the caption. The inserted Fourier plots show that, under this parameter settings, the Universe evolves as an OSS in the radiation era, and as a CAS in the matter era. In the radiation era, $\Omega_\phi\approx4\%$ as we expected. However, this parameter setting is difficult to explain the cosmic late-time acceleration in the matter era. An apparent difficulty is that $\Omega_\phi$ cannot reach $70\%$, i.e., the current observed dark energy relative density \cite{Aghanim2020.AstronAstrophys.641.A6}. Increasing $\lambda_1$ will exacerbate this problem. Increasing $\alpha$ can increase the maximum value of $\Omega_\phi$ in the matter era, but also make the cosmic evolution in the radiation era enter the CAS\footnote{In current parameter settings, $f_\textsc{oss}/2$ component already appears in the radiation era [see the left inset in Fig. \ref{fig:02} (a)]. Further parameter changes will cause the system to quickly enter the chaotic phase.}. Therefore, if $\lambda_1\gtrsim20$, then the OSS in the radiation era may be incompatible with the desired acceleration in the matter era.

What we need is a small $\Omega_\phi$ during BBN. To achieve this, in addition to increasing $\lambda_1$, we can also adjust the evolution of the Universe in the radiation era to CAS. Note that $\Omega_\phi$ can be very close to $0$ and $1$ in a CAS, but not in an OSS \cite{Tian2020.PRD.102.063509}. Technically, decreasing $\lambda_1$ or increasing $\alpha$ can bring the system into the CAS phase. In Fig. \ref{fig:02} (b), we plot such a scenario. A key setting here is $\lambda_1=10$ and other settings can be found in the caption. The model might survive the BBN constraints if $\Omega_\phi\lesssim4\%$ during the main BBN era. In addition, this parameter setting should also be able to explain the cosmic late-time acceleration (mainly $\Omega_\phi\approx70\%$ and the EOS $w_\phi\approx-1$) as shown in the right part of Fig. \ref{fig:02} (b) and the discussions in \cite{Tian2020.PRD.101.063531,Tian2020.PRD.102.063509}. Note that, in the matter era of Fig. \ref{fig:02} (b), we tend to emphasize the existence of the $\Omega_\phi=70\%$ points, rather than requiring it to occur exactly at today, i.e., $N=N_{\rm eq}+8.13$ \cite{Tian2020.PRD.102.063509}. Once the point exists, we can adjust the initial conditions so that it occurs right now, and this does not require fine-tuning \cite{Tian2020.PRD.102.063509}.

\begin{figure*}[!t]
  \centering
  \includegraphics[width=0.49\linewidth]{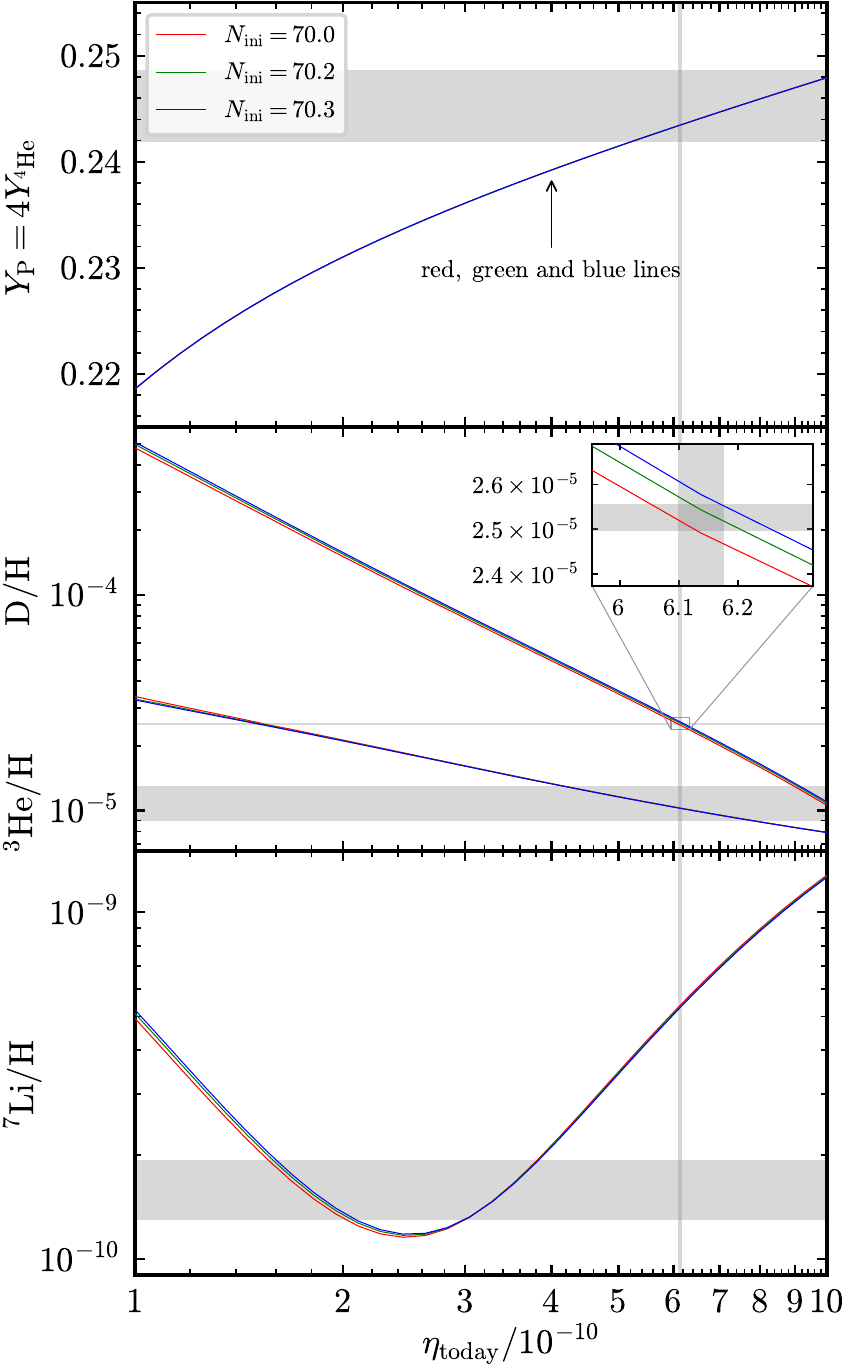}
  \includegraphics[width=0.49\linewidth]{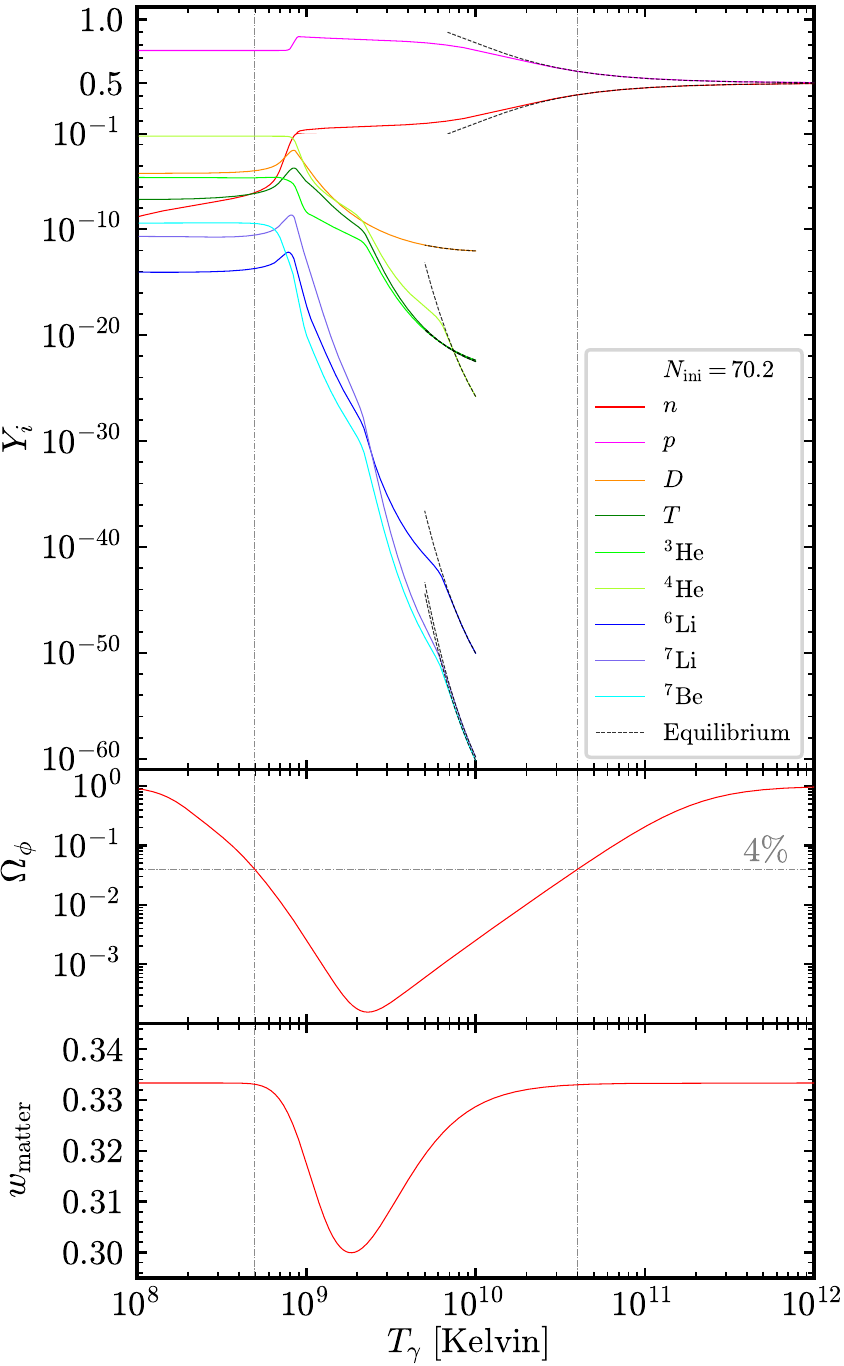}
  \caption{The BBN results for the stepwise scalar field model with the auxiliary parameter $N_{\rm ini}=70.0,\,70.2,\,70.3$. The model parameters are the same as in Fig. \ref{fig:02} (b). In the right part, the vertical dash-dotted line marks the point where $\Omega_\phi$ cross $4\%$. Others are the same as in Fig. \ref{fig:01}.}
  \label{fig:03}
\end{figure*}

\begin{figure*}[!t]
  \centering
  \includegraphics[width=0.49\linewidth]{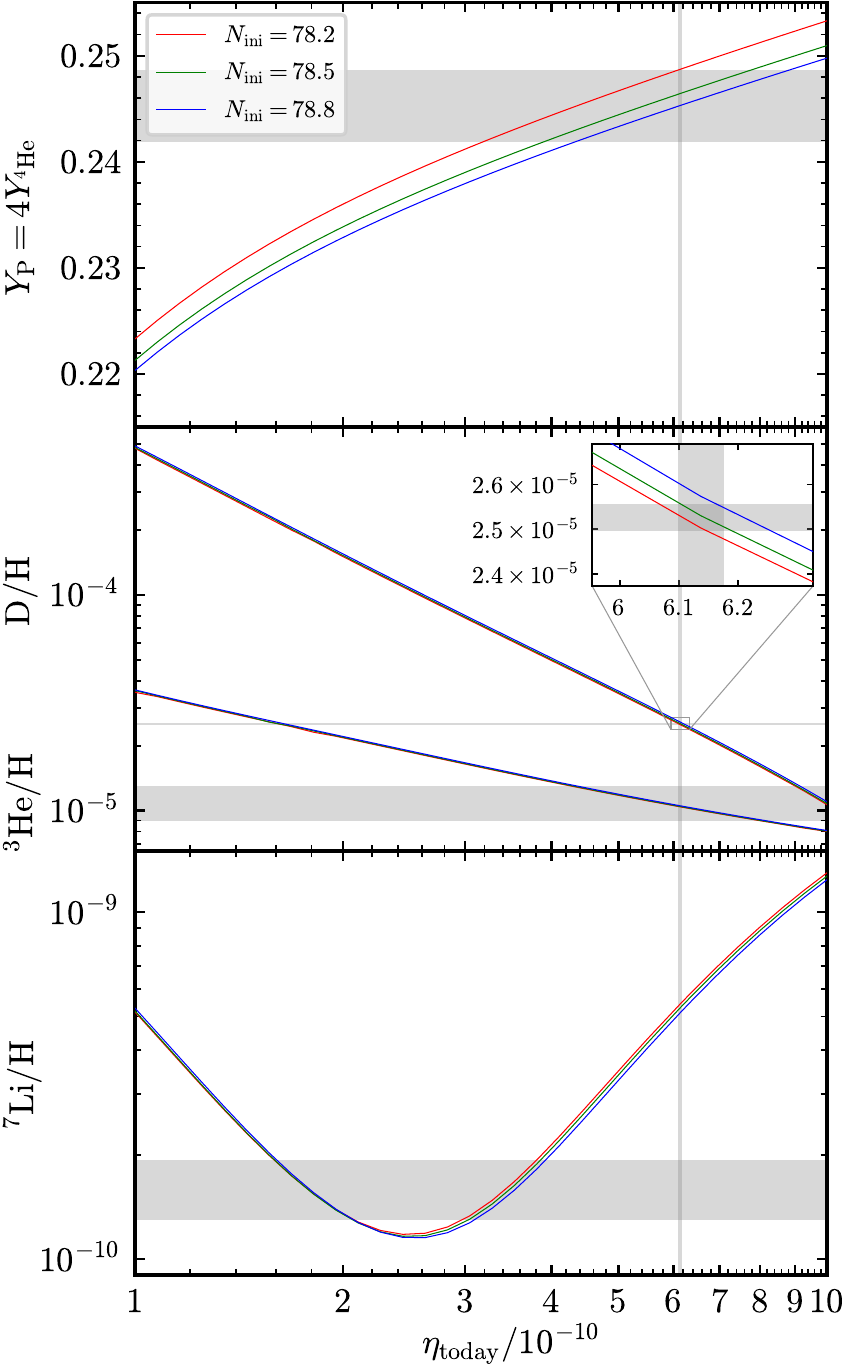}
  \includegraphics[width=0.49\linewidth]{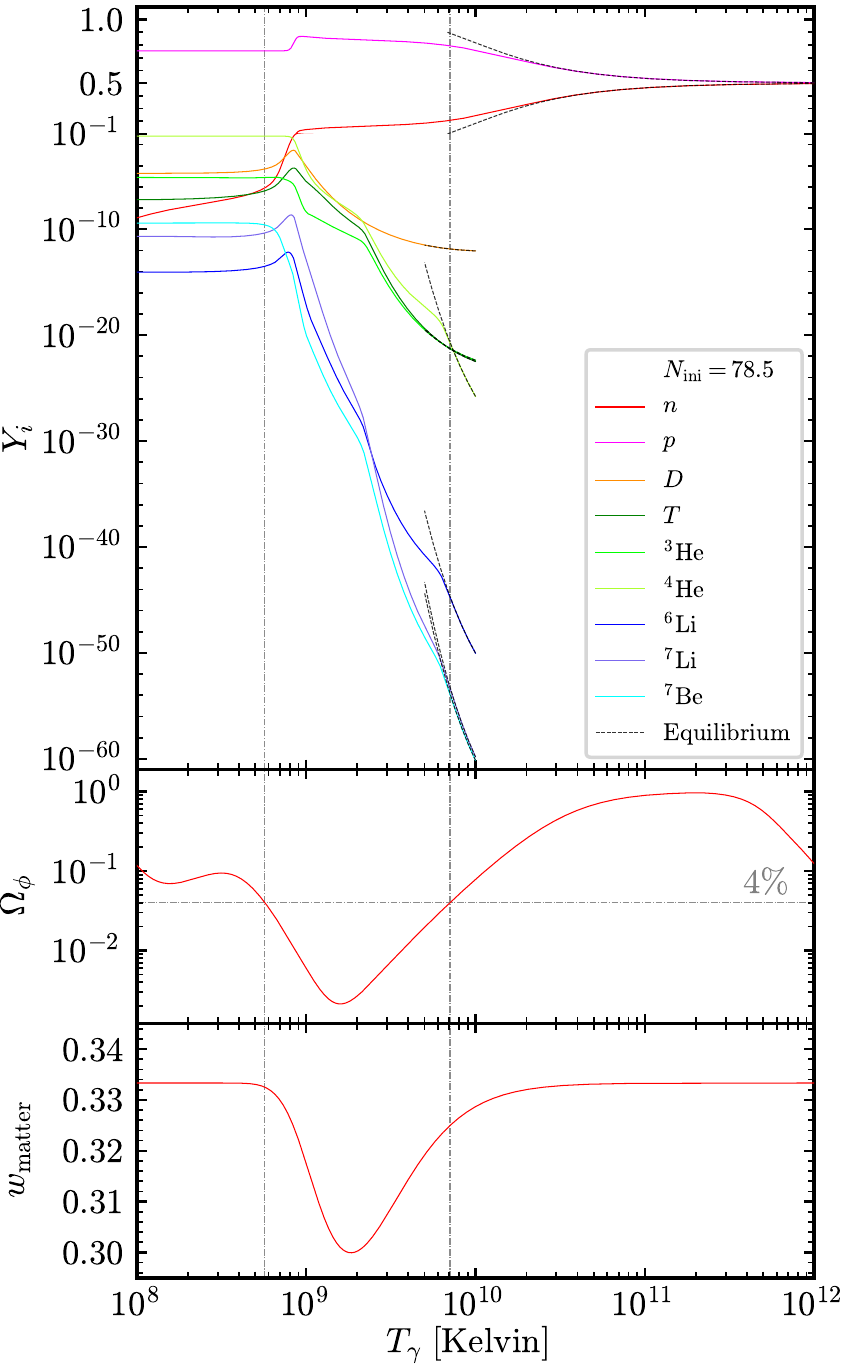}
  \caption{Same as Fig. \ref{fig:03} but with $N_{\rm ini}=78.2,\,78.5,\,78.8$.}
  \label{fig:04}
\end{figure*}

For the same parameter settings as Fig. \ref{fig:02} (b), we perform detailed BBN calculations with \texttt{BBNLab}. The results are shown in Figs. \ref{fig:03} and \ref{fig:04}. In order to set the initial conditions of the scalar field at $T_\gamma=10^{12}\,{\rm K}$ (see the discussions in Sec. \ref{sec:02}), we need an auxiliary parameter $N_{\rm ini}$, which is labeled in the figures. The colored shaded region in Fig. \ref{fig:02} (b) roughly corresponds to the BBN era of the $N_{\rm ini}=70.2$ case in Fig. \ref{fig:03}. Note that this correspondence is not strict, since $w_{\rm matter}$ (i.e., $w_{\rm m}$ in \cite{Tian2020.PRD.102.063509}) remains almost constant in the former case, while $w_{\rm matter}$ is obviously time-varying in the latter case. Looking at the $Y_{\rm P}$ and ${\rm D}/{\rm H}$ results in Figs. \ref{fig:03} and \ref{fig:04}, we conclude that there do exist viable parameter space in the stepwise scalar field model to explain the BBN observations. In particular, the case of $N_{\rm ini}=70.2$ in Fig. \ref{fig:03} or $N_{\rm ini}=78.5$ in Fig. \ref{fig:04} could be used to solve the possible deuterium problem \cite{Pitrou2021.MNRAS.502.2474}.

For the cases of $N_{\rm ini}=70.2$ and $N_{\rm ini}=78.5$, the evolutions of $\{Y_i,\Omega_\phi,w_{\rm matter}\}$ as a function of $T_\gamma$ are depicted in the corresponding right parts. There is one difference between the two cases worth mentioning: the $Y_{\rm P}$ predictions are different. Compared with the standard model, the case of $N_{\rm ini}=70.2$ does not change $Y_{\rm P}$ while the case of $N_{\rm ini}=78.5$ increases $Y_{\rm P}$. This difference should be due to the different values of $\Omega_\phi$ when protons and neutrons are out of equilibrium (see the second vertical dash-dotted line in the right parts of the figures). If $\Omega_\phi$ drops below $4\%$ before protons and neutrons fall out of equilibrium, then the scalar field would have no observable effect on the isotope evolutions in this period. In the case of $N_{\rm ini}=70.2$, the scalar field affects the isotope evolutions only near the end of BBN, and only has an observable effect on the deuterium abundance. This discussion also provides a simple mechanism for solving the deuterium problem: very early dark energy appears near the end of BBN. Essentially, this mechanism is the core of solving the deuterium problem with the stepwise scalar field. Finally, we admit that no solution of the lithium problem was found in the stepwise scalar field model.

\section{Conclusions}\label{sec:04}
The stepwise scalar field model was constructed to solve the cosmological coincidence problem \cite{Dodelson2000.PRL.85.5276,Tian2020.PRD.101.063531}. A key idea of this model is that the cosmic expansion has been dominated by the scalar field many times, which indicates that a non-negligible scalar field could have arisen at any stage in the early Universe. If it appears during BBN era, then the abundances of light elements will be affected due to the increment of the Hubble expansion rate. In this paper, we analyze the BBN consequences of the stepwise scalar field proposed in \cite{Tian2020.PRD.101.063531}. We provide here a public \texttt{Matlab} program, \texttt{BBNLab}, which enables the BBN calculation in the stepwise scalar field model. We find that BBN provides a strong upper limit $(\sim4\%)$ to the nearly constant $\Omega_\phi$ (see Fig. \ref{fig:01}). However, in the stepwise scalar field model, $\Omega_\phi$ can vary over several orders of magnitude, and can remain less than $4\%$ over a wide period of time. Based on this, we show examples where the stepwise scalar field can survive the BBN constraints. The scalar field can be completely hidden in BBN era, which restores the standard BBN results (see the case of $N_{\rm ini}=70.0$ in Fig. \ref{fig:03} or $N_{\rm ini}=78.2$ in Fig. \ref{fig:04}). A more interesting example is that the very early dark energy that appears near the end of BBN can be used to solve the possible deuterium problem \cite{Pitrou2021.MNRAS.502.2474} (see the case of $N_{\rm ini}=70.2$ in Fig. \ref{fig:03}).

In the stepwise scalar field model, the evolution of the Universe can be divided into two categories: OSS and CAS \cite{Tian2020.PRD.102.063509}. In order to facilitate the cosmological parameter constraints, we argued that the Universe should evolve as OSS in the radiation era \cite{Tian2020.PRD.102.063509}. This is a subjective argument, not a result of observational constraints. However, the above $4\%$ upper limit, together with the constraints from the cosmic late-time acceleration, rule out the possibility of OSS in the radiation era. The examples provided in this paper that not only survive the BBN constraints, but also explain the cosmic late-time acceleration, are all belong to the category of CAS. Therefore, we conclude that the Universe should evolve as a CAS in the radiation era in the stepwise scalar field model. This does not rule out the model, but brings difficulties to the global cosmological parameter constraints: there may be many local maxima in the posterior distribution \cite{Tian2020.PRD.101.063531}. How to quantitatively describe such a posterior distribution is a major task for our future work.

What kinds of observations could verify or rule out the CAS scenario for the radiation era? Given the above discussed difficulty of cosmological parameter constraints, it is important to look for direct observations. Electromagnetic observations seem impossible due to the scattering of photons and free electrons. However, gravitational waves and dark matter observations might be possible. The stochastic gravitational wave background can be used to infer the cosmic expansion history before BBN \cite{Boyle2008.PRD.77.063504,Cui2018.PRD.97.123505,DEramo2019.PhysRevResearch.1.013010}. The possible axion detection provides an auxiliary probe of the early Universe \cite{Ramberg2019.PRD.99.123513}. Future multimessenger astronomy can directly probe the early Universe and test the CAS scenario in our model.

\section*{Acknowledgements}
This work was supported by the Initiative Postdocs Supporting Program under Grant No. BX20200065 and China Postdoctoral Science Foundation under Grant No. 2021M700481.

\appendix
\section{Special functions}\label{App:A}
Here we summarize the special functions involved in the BBN calculation:
\begin{align}
  M(x)&=\frac{1}{x}\left[\frac{3}{4}K_3(x)+\frac{1}{4}K_1(x)\right], \\
  M'(x)&=-(\frac{24}{x^4}+\frac{5}{x^2})K_1(x)-(\frac{12}{x^3}+\frac{1}{x})K_0(x), \\
  L(x)&=\frac{1}{x}K_2(x), \\
  L'(x)&=-(\frac{1}{x}+\frac{6}{x^3})K_1(x)-\frac{3}{x^2}K_0(x),
\end{align}
where the second kind modified Bessel function \cite{Chandrasekhar1939.book}
\begin{equation}
  K_\nu(x)=\int_0^{+\infty}e^{-x\cosh\theta}\cosh(\nu\theta)\,\dx\theta.
\end{equation}
In \texttt{Matlab}, $K_\nu(x)$ is calculated by $\texttt{besselk}(\nu,x)$.

\end{document}